\documentclass[twocolumn,superscriptaddress,preprintnumbers,amsmath,amssymb,floatfix]{revtex4}
\pdfoutput=1
\usepackage{graphicx} 
\usepackage{dcolumn} 
\usepackage{bm}
\usepackage{color}
\usepackage{hyperref}
\hypersetup{
    colorlinks=true,
    linkcolor=[RGB]{0,0,139}, %
    filecolor=magenta,
    urlcolor=cyan,
    citecolor=[RGB]{0,0,139}, %
    pdfborder={0 0 0} %
}

\usepackage{graphicx}
\usepackage{amsmath}
\usepackage{setspace}
\usepackage{appendix}
\usepackage[T1]{fontenc}
\usepackage{textcomp}
\usepackage{txfonts}
\usepackage{float}
\usepackage{booktabs}

\newcommand{\bmr}{\textbf{\textit{r}}}
\usepackage{ulem}
\usepackage{xcolor}
\DeclareRobustCommand{\erase}{\bgroup\markoverwith{\textcolor{red}{\rule[.5ex]{2pt}{0.4pt}}}\ULon}

\begin{document}

\title{The quenching of the axial-vector coupling constant $g_A$ in $\beta$-decay: joint effects from chiral two-body currents and many-body correlations}

\author{Bin-Lei Wang}
\affiliation{Key Laboratory of Beam Technology of Ministry
of Education, School of Physics and Astronomy, Beijing
Normal University, Beijing 100875, China}

\author{Wan-Li Lv}
\affiliation{School of Nuclear Science and Technology, Lanzhou University, Lanzhou 730000, China}
\affiliation{Frontiers Science Center for Rare isotopes, Lanzhou University, Lanzhou 730000, China}

\author{Li-Gang Cao}
\email{caolg@bnu.edu.cn}
\affiliation{Key Laboratory of Beam Technology of Ministry
of Education, School of Physics and Astronomy, Beijing
Normal University, Beijing 100875, China}
\affiliation{Key Laboratory of Beam Technology of Ministry
of Education, Institute of Radiation Technology, Beijing
Academy of Science and Technology, Beijing 100875, China}

\author{Yi-Fei Niu}
\email{niuyf@lzu.edu.cn}
\affiliation{School of Nuclear Science and Technology, Lanzhou University, Lanzhou 730000, China}
\affiliation{Frontiers Science Center for Rare isotopes, Lanzhou University, Lanzhou 730000, China}

\author{Gianluca Col\`o}
\email{Gianluca.Colo@mi.infn.it}
\affiliation{Dipartimento di Fisica, Universit\`a degli Studi di Milano, and INFN, Sezione di Milano, Via Celoria 16, 20133 Milano, Italy}

\author{Hiroyuki Sagawa}
\affiliation{RIKEN Nishina Center, Wako 351-0198, Japan}
\affiliation{Center for Mathematics and Physics, University of Aizu, Aizu-Wakamatsu 965-8560, Japan}
\affiliation{Institute of Theoretical Physics, Chinese Academy of Sciences, Beijing 100190, China}

\author{Feng-Shou Zhang}
\affiliation{Key Laboratory of Beam Technology of Ministry
of Education, School of Physics and Astronomy, Beijing
Normal University, Beijing 100875, China}
\affiliation{Key Laboratory of Beam Technology of Ministry
of Education, Institute of Radiation Technology, Beijing
Academy of Science and Technology, Beijing 100875, China}
\affiliation{Center of Theoretical Nuclear Physics, National Laboratory of Heavy Ion Accelerator of Lanzhou, Lanzhou 730000, China}

\date{\today}

\begin{abstract}
In nuclear $\beta$-decay calculations, the axial-vector coupling constant $g_A \approx 1.27$ usually needs to be quenched phenomenologically by a factor $q~\approx$ 0.75 to reproduce {the Gamow-Teller (GT) transition strengths}. We propose a novel approach to quench the GT {strength} of $\beta$-decay within the microscopic random phase approximation (RPA) plus particle-vibration coupling (PVC) approach, incorporating the contributions of two-body currents (TBC) derived from chiral effective field theory ($\chi$EFT). Self-consistent RPA+PVC calculations are performed in three doubly magic nuclei, $^{56}$Ni, $^{100}$Sn, and $^{132}$Sn,  with various Skyrme energy density functionals, and the effect of
TBC is evaluated by using the obtained many-body wavefunctions. A combined effects of  the many-body correlations introduced by PVC  and chiral TBC  quench the GT strength  and  reproduce quantitatively experimental data without any additional adjustments.
The extracted quenching factors $q$ by the present microscopic model lie in the range $\approx$ 0.73--0.80, which is quite close to  the commonly adopted empirical value of $q \approx 0.75$.

\end{abstract}

\maketitle



\textit{Introduction}-- Nuclear $\beta$-decay plays a crucial role in nuclear physics, astrophysics and particle physics. The study of $\beta$-decay helps us to understand the {nucleon-nucleon} interactions in the spin and spin-isospin channels~\cite{RevModPhys.64.491,PhysRevC.65.054322,PhysRevC.76.044307,PhysRevC.89.044311}.  The {$\beta$-decay half-lives set the timescale for rapid neutron-capture process that is responsible for the synthesis of heavy elements in the universe}~\cite{Kajino:2019abv,RevModPhys.93.015002}.
Searches for neutrinoless double-$\beta$ decay provide the chance to test the standard model and to probe the basic properties of neutrinos~\cite{Bilenky15,RevModPhys.80.481}.

The Gamow-Teller (GT) transitions represent the dominant nuclear transition mode in $\beta$-decay. However, the GT transition strengths calculated using various microscopic models---such as the proton-neutron quasiparticle random phase approximation (pn-QRPA)~\cite{SUHONEN198891,10.3389/fphy.2019.00030,PhysRevC.86.015809}  and the shell model~\cite{WILKINSON1973470,PhysRevC.7.930,WOS:A1988R251900002}---tend to systematically overestimate the experimental data. Specifically, the low-energy GT strength for a single transition relevant to $\beta$-decay is larger than the value deduced from the experimental $\log ft$ value. Empirically, this discrepancy is often resolved by introducing a phenomenological quenching factor $q$ \cite{TOWNER1987263, WOS:A1985AXS5900001, PhysRevC.47.163, PhysRevC.53.R2602,PhysRevC.100.054324,PhysRevC.104.054318}, which effectively reduces the axial-vector coupling constant from its free-nucleon value $g_A \approx 1.27$ \cite{PhysRevLett.122.242501} in the calculation of $\beta$-decay nuclear matrix elements. This longstanding discrepancy is referred to as the problem of $g_A$ quenching in nuclear weak decay processes.

The origin of the $g_A$ quenching is generally attributed to two main sources. The first source is the lack of sufficient many-body correlations in the nuclear wavefunctions. For instance, the shell model typically employs a truncated model space, while the RPA model restricts the model space to the one particle--one hole ($1p$-$1h$) configurations in a large energy range. Regarding improvements, shell model studies have demonstrated that enlarging the model space is essential for a better description of the GT strength \cite{PhysRevLett.110.222502,PhysRevC.48.1677}. Alternatively, the inclusion of $2p$-$2h$ configurations beyond the standard RPA framework---as implemented in particle-vibration coupling (PVC) model \cite{PhysRevC.90.054328,LITVINOVA2014307,WOS:000391354400001,PhysRevC.98.051301} and the subtracted second RPA (SSRPA) model \cite{PhysRevLett.125.212501,PhysRevC.105.014321,PhysRevC.106.014319}---has been proven crucial for addressing the quenching problem. Notably, recent SSRPA calculations have shown that incorporating $2p$-$2h$ configurations improve significantly the description of both the summed GT strength up to $20$ MeV in $^{48}$Ca and the $\beta$-decay half-lives using a bare $g_A$ in $^{78}$Ni \cite{PhysRevLett.125.212501}.

The second source comes from the two-body currents (TBC) in the electroweak decay operator, which arise from the intrinsic structure of non point-like nucleons. In Ref.~\cite{PhysRevLett.107.062501}, it is shown that the inclusion of chiral TBC provides important contributions to the quenching of low-momentum-transfer GT transitions in the shell model calculations.  Based on the QRPA approaches, the effects of TBC on GT strength, $\beta$-decay rates, and the nuclear matrix elements (NMEs) of neutrinoless double-$\beta$ decay have been systematically investigated in Refs.~\cite{PhysRevC.89.064308,PhysRevC.105.034349}. These studies indicate that TBC tends to quench both the summed GT strength and the decay rates. Recently, research works based on {\it ab initio} calculations have been devoted to analyze the $g_A$ quenching by including the TBC~ \cite{WOS:000466716100012,PhysRevLett.113.262504,xjv9-t6sn} in the processes of $\beta$-decay and neutrinoless double-$\beta$ decay. It has been shown that the inclusion of TBC in such calculations can successfully explain the $g_A$ quenching puzzle. However, it is also shown that the inclusion of TBC does not always quench the GT strength, but  enhances the GT strength in some light nuclei~\cite{WOS:000466716100012}.

Although the general picture seems reasonable, the relative roles of many-body correlations and two-body currents in explaining the quenching problem are not well understood. A work based on the shell model approach, employing many-body perturbation theory to derive the GT operator and TBC, is an attempt in this direction~\cite{PhysRevC.109.014301}. Nevertheless, further efforts are required to fully address this question. The PVC approach includes more complex configurations beyond RPA such as $1p$-$1h$ coupled with phonons to incorporate many-body correlations. Therefore, it provides us with an ideal tool to study the effects of both many-body correlations and TBC on the GT strength in $\beta$ decay. In this letter, we present for the first time the  study of GT strength of $\beta$-decay within RPA + PVC model, incorporating the contributions of TBC from $\chi $EFT \cite{PhysRevC.67.055206,PhysRevLett.103.102502}. The RPA+PVC calculations are performed
in the doubly magic nuclei $^{56}$Ni, $^{100}$Sn, and $^{132}$Sn,  by using various Skyrme energy density functionals (EDFs). Our aim is to quantify the quenching effect induced by {both} TBC and {complex many-body configurations} to examine whether such a mechanism provides a unified microscopic explanation for the long-standing quenching problem.

\textit{Theoretical framework}-- Here the basic formulas will be
presented, whereas more details can be found in the Supplemental Material~\cite{Wangbl}. With the long-wavelength approximation, the transition operator in nuclear weak (single-$\beta$) decay corresponds to the nuclear weak currents comprising the vector and axial-vector components. When considering only the axial-vector currents in the zero-momentum-transfer limit, the one-body current (OBC) operator 
is expressed as \cite{PhysRevC.105.034349},
\vspace{-2mm}
\begin{eqnarray}
    \mathbb{J}_{1 b}^A(\boldsymbol{x}) = -g_A\sum_{i} \bm{\sigma}_i \bm{\tau}_i^{\pm} \delta(\bm{x} - \bmr_i). \label{1bc-operator}
\end{eqnarray}
where $\bm\sigma$ denotes the Pauli spin operator, and $\bm{\tau}^{\pm}$ is the isospin raising or lowering operator.

For the chiral TBC, we adopt the formulas in Refs.~\cite{PhysRevC.67.055206,PhysRevLett.103.102502}. Using Fourier transforming, the axial TBC operator in momentum-space can be transformed into coordinate-space, which is expressed as \cite{PhysRevC.67.055206,PhysRevC.98.031301}:
\begin{widetext}
\vspace{-8mm}
\begin{eqnarray}
  \mathbb{J}_{2b}^A(\boldsymbol{x}) &=& \sum_{k\,<\,l} \frac{2\bar{c}_3 g_A}{m_N F_{\pi}^2} \left\{ m_{\pi}^2 \left[ \left( \frac{\bm{\sigma}_l}{3} - \bm{\sigma}_l \cdot \hat{\bmr}\, \hat{\bmr} \right) Y_2(r) - \frac{\bm{\sigma}_l}{3} Y_0(r) \right] + \frac{\bm{\sigma}_l}{3} \delta(\bmr) \right\} \bm{\tau}_l^{\pm} \delta(\bm{x} - \bmr_k) + (k \leftrightarrow l) \notag\\
  &+& \left( \bar{c}_4 + \frac{1}{4} \right) \frac{g_A}{2m_N F_{\pi}^2} \left\{ m_{\pi}^2 \left[ \left( \frac{\bm{\sigma}_\times}{3} - \bm{\sigma}_k \times \hat{\bmr}\, \bm{\sigma}_l \cdot \hat{\bmr} \right) Y_2(r) - \frac{\bm{\sigma}_\times}{3} Y_0(r) \right] + \frac{\bm{\sigma}_\times}{3} \delta(\bmr) \right\} \bm{\tau}_\times^{\pm} \delta(\bm{x} - \bmr_k) + (k \leftrightarrow l) \notag \\
  &+& \frac{ig_A}{8m_N F_{\pi}^2} \,\bm{\tau}_\times^{\pm}\, \left[ \bm{\sigma}_l \cdot \hat{\bmr}\, \hat{\bmr} \, m_{\pi}^2 \left( 1 + \frac{2}{m_{\pi}r} + \frac{2}{m_{\pi}^2 r^2} \right) Y_0(r) + \frac{\bm{\sigma}_l \cdot \hat{\bmr}\,\hat{\bmr}}{r^2} Y_1(r) - \frac{\bm{\sigma}_l}{r^2} Y_1(r) \right] \delta(\bm{x} - \bmr_k) + (k \leftrightarrow l) \notag \\
  &-& \frac{g_A}{4m_N F_{\pi}^2} \left[ 2 \bar{d}_1 \left(\bm{\sigma}_k \bm{\tau}_k^{\pm} + \bm{\sigma}_l \bm{\tau}_l^{\pm}\right) + \bar{d}_2 \bm{\sigma}_\times \bm{\tau}_\times^{\pm} \right] \delta(\bmr) \delta(\bm{x} - \bmr_k), \label{TBC-operator}
\end{eqnarray}
\end{widetext}
where $F_\pi = 92.4$ MeV is the pion decay constant, $m_\pi$ is the pion mass, $r = r_k - r_l$, and $\hat{\bmr} \equiv \frac{\bmr}{r}$. The Yukawa functions are $Y_0(r) = \frac{e^{-m_\pi r}}{4\pi r}$, $Y_1(r) = -r\frac{\partial}{\partial r} Y_0(r)$ and $Y_2(r) = \frac{1}{m_\pi^2} r \frac{\partial}{\partial r} \frac{1}{r} \frac{\partial}{\partial r} Y_0(r)$. The vector products of spin and isospin operators are defined as $\bm\sigma_\times = \bm\sigma_k \times \bm\sigma_l$ and $\bm{\tau}^{\pm}_\times = (\bm{\tau}_k \times \bm{\tau}_l)^{\pm}$ \cite{PhysRevC.67.055206}. The dimensionless low-energy constants (LECs) are $\bar c_3, \bar c_4, \bar d_1, \bar d_2$ with the definition $\bar c_{3, 4} = m_N c_{3, 4}$, $\bar d_{1, 2} = \frac{m_N F_\pi^2}{g_A} d_{1, 2}$ and $\bar c_{D} \equiv \bar d_1 + 2 \bar d_2$~\cite{PhysRevC.67.055206}. Furthermore, the first three terms of $\mathbb{J}_{2b}^A(\boldsymbol{x})$  in Eq. (\ref{TBC-operator}) correspond to the finite-range components associated with the $\bar{c}_3$, $\bar{c}_4$, and {momentum-dependent} term, respectively, while the last line represents the short-range $\bar{c}_D$ term.

In the present study,  {GT matrix elements} with both the OBC and chiral TBC are evaluated using either the RPA or RPA + PVC wavefunctions, based on the following expression:
\begin{widetext}
\vspace{-6mm}
\begin{eqnarray}
  \left\langle 1^+_\omega \left| \mathbb{M}_{1b}^\lambda+\mathbb{M}_{2b}^\lambda \right| 0^+\right\rangle
  \overset{\rm QBA}{\approx}
  \left\langle {\rm HF}\left|\left[ \mathcal{O}_{1\omega},\mathbb{M}_{1b}^\lambda \right]\right|{\rm HF} \right\rangle +  \left\langle {\rm HF}\left|\left[ \mathcal{O}_{2\omega},\mathbb{M}_{1b}^\lambda \right]\right|{\rm HF} \right\rangle +  \left\langle {\rm HF}\left|\left[ \mathcal{O}_{1\omega},\mathbb{M}_{2b}^\lambda \right]\right|{\rm HF} \right\rangle +  \left\langle {\rm HF}\left|\left[ \mathcal{O}_{2\omega},\mathbb{M}_{2b}^\lambda \right]\right|{\rm HF} \right\rangle, \label{NMEs}
\end{eqnarray}
\end{widetext}
where $\mathbb{M}_{1b}^\lambda$ and $\mathbb{M}_{2b}^\lambda$ represent the GT transition operators obtained by integrating the current operator $\mathbb{J}_{1b}^A$ and $\mathbb{J}_{2b}^A$ over $\bm{x}$ under the long-wavelength approximation, namely $\mathbb{M}^\lambda=\int d\bm{x} e^{-i\bm{q}\cdot\bm{x}}\mathbb{J}_{\mu}(\bm{x})\approx \int d\bm{x} \mathbb{J}_{\mu}(\bm{x})$. $\mathcal{O}_{1\omega}$ and $\mathcal{O}_{2\omega}$ represent the $1p$-$1h$ component and $1p$-$1h$ $\otimes$ phonon component of the PVC operator~\cite{PhysRevC.110.064317}, and QBA stands for quasi-boson approximation \cite{2007From}. In Eq. (\ref{NMEs}), the RPA model contains only the first and third {matrix elements} because the RPA wavefunctions lack the $1p$-$1h$ $\otimes$ phonon component $\mathcal{O}_{2\omega}$, while the RPA + PVC model contains all {matrix elements}. Since the operator $\mathcal{O}_{2\omega}$ contains higher-order nucleon correlations through the forward-going amplitudes $X_{\pi \bar{\nu}n}$ and backward-going amplitudes $Y_{\nu \bar{\pi}n}$, it becomes necessary to solve the RPA + PVC equation presented below,

\begin{widetext}
\vspace{-5mm}
\begin{eqnarray}
  \left(\begin{array}{cccc}
        A_{\pi \bar{\nu},\pi^\prime \bar{\nu}^\prime} & B_{\pi\bar{\nu},\nu^\prime \bar{\pi}^\prime} & A_{\pi\bar{\nu},\pi^\prime \bar{\nu}^\prime n^\prime} & 0 \\
        -B^\ast_{\nu\bar{\pi},\pi^\prime \bar{\nu}^\prime} & -A^\ast_{\nu\bar{\pi},\nu^\prime \bar{\pi}^\prime} & 0 & -A^\ast_{\nu\bar{\pi},\nu^\prime \bar{\pi}^\prime n^\prime} \\
        A_{\pi\bar{\nu}n,\pi^\prime \bar{\nu}^\prime} & 0 & A_{\pi\bar{\nu}n,\pi^\prime \bar{\nu}^\prime n^\prime} & 0 \\
        0 & -A^\ast_{\nu \bar{\pi} n,\nu^\prime \bar{\pi}^\prime} & 0 & -A^\ast_{\nu \bar{\pi} n,\nu^\prime \bar{\pi}^\prime n^\prime}
    \end{array}\right)
    \left(\begin{array}{ccc}
        X^\omega_{\pi^\prime \bar{\nu}^\prime}   \\
        Y^\omega_{\nu^\prime \bar{\pi}^\prime}   \\
        X^\omega_{\pi^\prime \bar{\nu}^\prime n^\prime}   \\
        Y^\omega_{\nu^\prime \bar{\pi}^\prime n^\prime}
    \end{array}\right)
    = E_\omega
        \left(\begin{array}{ccc}
        X^\omega_{\pi \bar{\nu}}   \\
        Y^\omega_{\nu \bar{\pi}}   \\
        X^\omega_{\pi \bar{\nu} n}   \\
        Y^\omega_{\nu \bar{\pi} n}
        \end{array}\right). \label{RPAPVC}
\end{eqnarray}
\end{widetext}

\begin{figure*}
\begin{center}
  \includegraphics[width=0.850\textwidth]{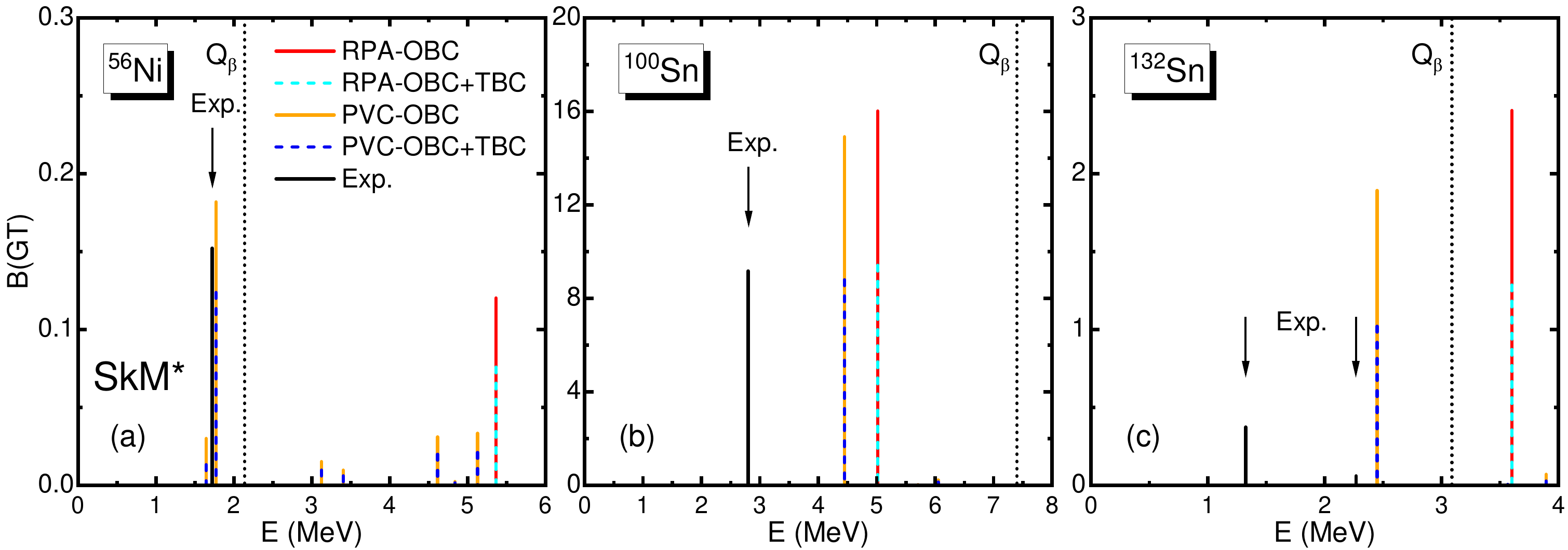}
  \caption{\label{energy}(Color online). Theoretical and experimental B(GT) distributions in the daughter nuclei of $^{56}$Ni, $^{100}$Sn, and $^{132}$Sn. The excitation energy is referred to the ground state of daughter nucleus.  The theoretical results are obtained within RPA and RPA + PVC models with the SkM* interaction. The black vertical dotted lines indicate the experimental values of $Q_{\beta}$ \cite{nndc}. The red and yellow solid lines represent the RPA and RPA + PVC results with OBC, respectively. The cyan and blue dashed lines represent the RPA and RPA + PVC results together with the TBC. In Fig. 1(a) the strengths above the $Q_{\beta}$ are the values obtained by multiplying 1/100 on its real values. Experimental data are showed with black solid lines~\cite{nndc}. }
\vspace{-5mm}
\end{center}
\end{figure*}

\textit{Results and discussion}-- The main results will be presented here, while for the detailed numerical checks we refer to the Supplemental Material~\cite{Wangbl}. For the low-energy constants $c_3$ and $c_4$, there exist several parameter sets in the literature:   the EM set ($(c_3,c_4)$ = $(-3.2,5.4)$)~\cite{PhysRevC.68.041001}, the RTD set ($(c_3,c_4)$ = $(-4.78,3.96)$)~\cite{PhysRevC.67.044001}, and the EGM set ($(c_3,c_4)$ = $(-3.4,3.4)$)~\cite{EPELBAUM2005362} (all values are in GeV$^{-1}$). We use the most commonly used choices; despite some spread, these parameters are relatively well-known as they are related to pion-nucleon phenomenology. In contrast, the contact term $\bar{c}_D$ is far less known and we follow several authors in setting it to zero, by commenting on the sensitivity of the results to the variation of this parameter that could span positive or negative values in the range $\vert \bar{c}_D \vert \approx 1-10$. Fig. \ref{energy} shows the theoretical and experimental B(GT) distributions of $^{56}$Ni, $^{100}$Sn, and $^{132}$Sn that highlight both the PVC and TBC effects on the GT transitions, respectively. In the figure, we display results of calculations performed within RPA and RPA + PVC models, using the SkM* interaction, and TBC is included using the RTD set with LECs $(c_3,c_4)$ = $(-4.78,3.96)$ GeV$^{-1}$ \cite{PhysRevC.67.044001} and $\bar{c}_D=0$. Two effects can be seen clearly: the first one is the  effect of many-body correlations in the wavefunctions of excited states provided by the RPA + PVC calculations, which shifts the energies of 1$^+$ states downwards and  reduce the B(GT) values; the second one is the effect from  two-body meson-exchange currents, which systematically reduces the calculated strengths obtained by  both RPA and RPA + PVC models.

In $^{56}$Ni [Fig.~\hyperref[energy]{\ref*{energy}(a)}], experiments observed only one $1^{+}$ state at 1.72 MeV below  the $Q_\beta$ value: this is the final state of the orbital electron capture (EC) process~\cite{PhysRevC.42.573}. There is no strength below $Q_{\beta}$ in the RPA calculation. The RPA + PVC calculation allows this process giving two states below $Q_{\beta}$, with energies  very close to the experimental data. In the RPA+PVC results, the summed B(GT) value of 0.212 given by the OBC overestimates the experimental value 0.152 \cite{PhysRevC.42.573}, while this is  reduced to 0.137 by the TBC.

In $^{100}$Sn [Fig.~\hyperref[energy]{\ref*{energy}(b)}], experiments observed a strong GT transition at E=2.80 MeV with B(GT) $\approx$ 9.167~\cite{WOS:000305466800032}; the decay is 90\% through $\beta^{+}$ and 10\% through EC process~\cite{nndc}. Both RPA and RPA + PVC predict the $1^{+}$ states within the $\beta$-decay window. The PVC effect brings GT states closer to their experimental values both in energy and strength. When including the TBC, the GT strengths become  B(GT)=9.412 by  RPA and B(GT)=8.930 by RPA + PVC, respectively,  showing excellent agreement with the experimental data.

In $^{132}$Sn [Fig.~\hyperref[energy]{\ref*{energy}(c)}], experiments observed two $1^{+}$ states at energies 1.33 MeV and 2.27 MeV with the summed GT strength B(GT)=0.43~\cite{PhysRevC.51.500}. The RPA calculation predicts the lowest state at 3.60 MeV, which is outside of the $\beta$-decay window ($E=3.09$ MeV). The lowest state ($E=2.45$ MeV) in the RPA + PVC calculation is below $Q_{\beta}$. The calculated B(GT) are 2.404 and 1.892, respectively, in RPA and RPA + PVC with OBC. Similarly to the cases of $^{56}$Ni and  $^{100}$Sn just discussed, the PVC shifts the energy downwards and lowers the GT strength. The B(GT)  values, 1.285 and 1.021, for the RPA and RPA + PVC models with the TBC turn out to be much better agreement with the experimental B(GT) value.

\begin{figure}
\begin{center}
  \includegraphics[width=0.40\textwidth]{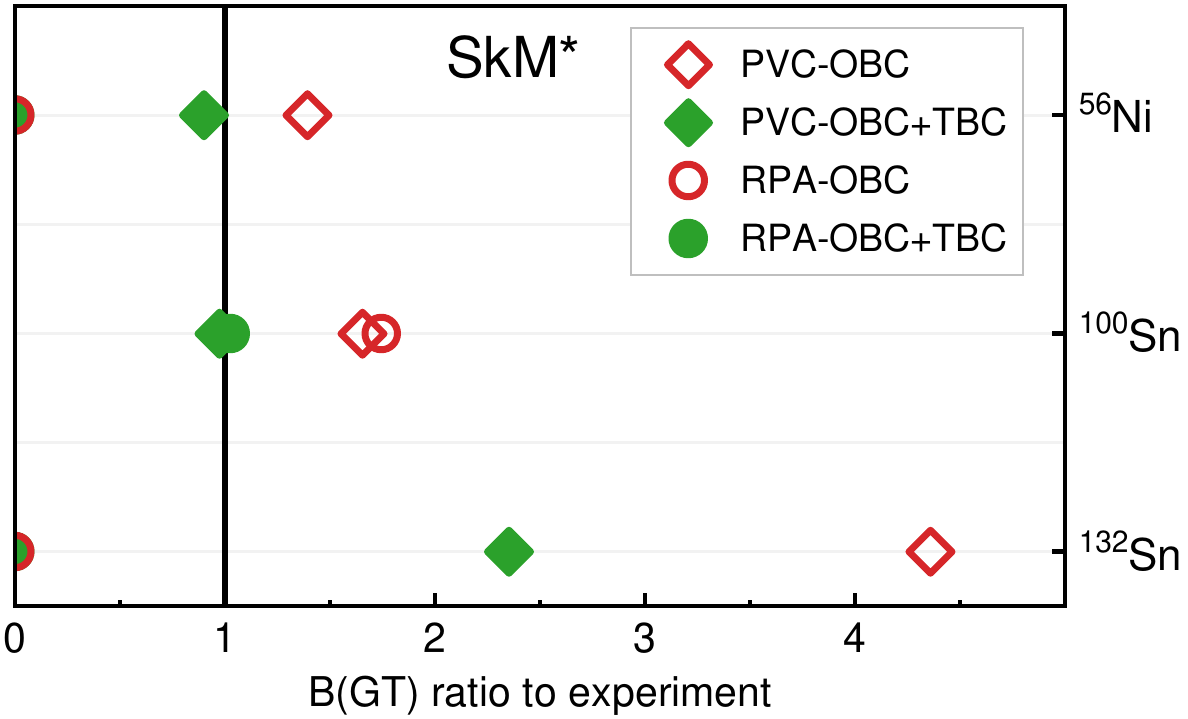}
  \caption{\label{ratio}(Color online). Ratios of theoretical to experimental B(GT) values  within $Q_\beta$ window for the doubly magic nuclei $^{56}$Ni, $^{100}$Sn, and $^{132}$Sn, computed within RPA and RPA + PVC models with the SkM* interaction. Open symbols denote the contributions from OBC only and filled symbols are the results including the TBC. The experimental data are taken from Refs.~\cite{PhysRevC.42.573,WOS:000305466800032,PhysRevC.51.500}.}
\vspace{-8mm}
\end{center}
\end{figure}

To clarify  the roles of PVC and chiral TBC, Fig. \ref{ratio} shows the ratios of theoretical to experimental B(GT) within the $Q_\beta$ window for $^{56}$Ni, $^{100}$Sn, and $^{132}$Sn. The results reveal that TBC systematically quench the B(GT) values obtained by PVC for all selected nuclei. For $^{100}$Sn, the calculated result nearly coincides with the experimental data. For $^{56}$Ni, the OBC overestimates the measured value, but the inclusion of TBC improves the agreement significantly. For $^{132}$Sn, the B(GT) given by OBC exceeds the experimental value by  a factor of 4.4, while the TBC leads again to a substantial improvement compared to the experimental data. In the RPA model, the nucleus $^{100}$Sn has GT strength below the $Q_\beta$ value and the $\beta$-decay is possible.  However, there is no strength in the $\beta$-decay window for $^{56}$Ni and $^{132}$Sn so that they cannot decay at the RPA level.

\begin{figure}
\begin{center}
  \includegraphics[width=0.400\textwidth]{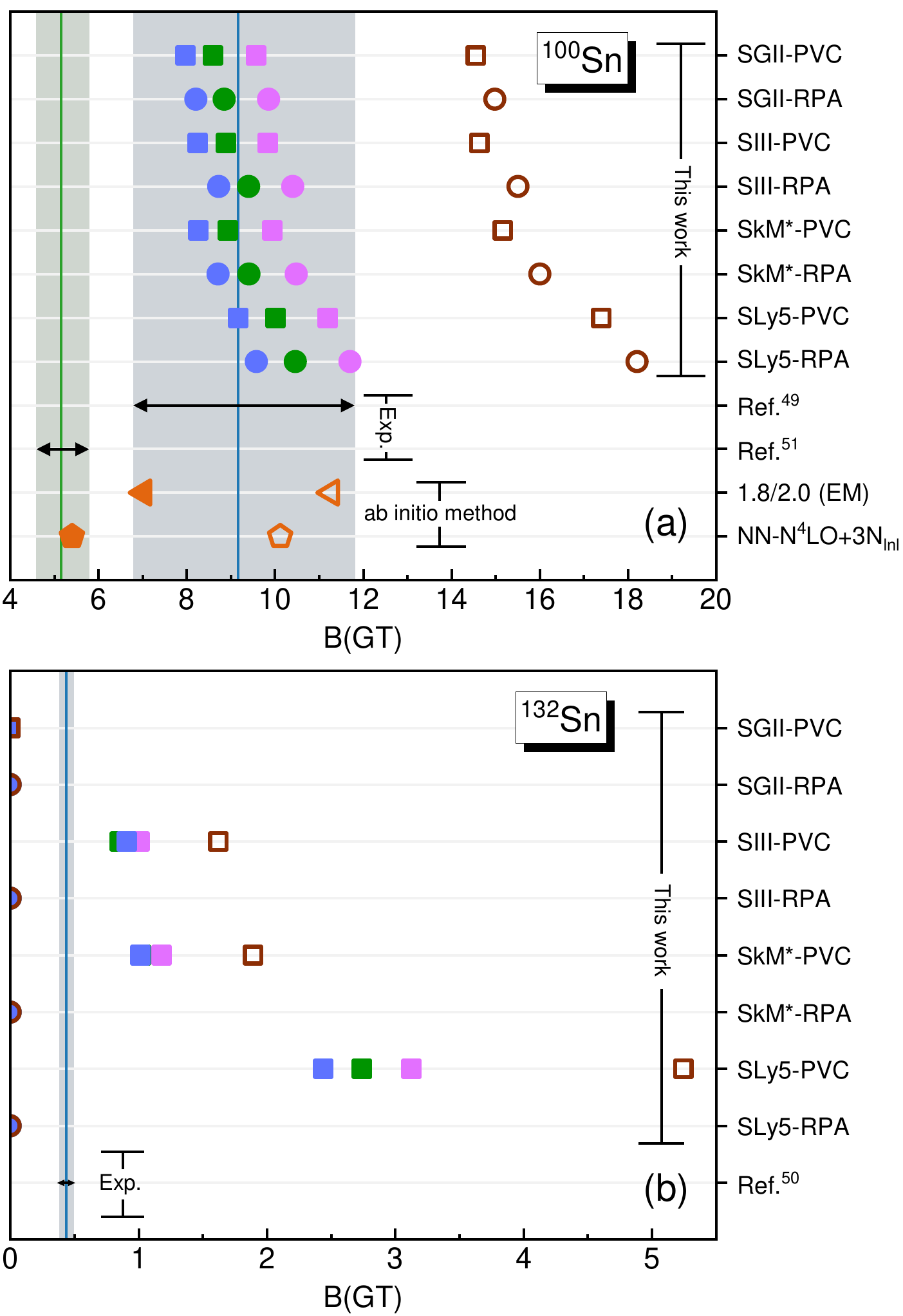}
  \caption{\label{100Sn-132Sn}(Color online). Comparison of the calculated B(GT) in $\beta$-decay with the experimental data~\cite{WOS:000305466800032,WOS:000282433500005,PhysRevC.51.500} and the {\it ab initio} calculations~\cite{WOS:000466716100012} for $^{100}$Sn and $^{132}$Sn. The results are obtained within RPA and RPA + PVC methods with different Skyrme interactions. Open symbols denote OBC results, filled symbols are the results obtained with OBC+TBC, and blue, green and purple symbols correspond to the results with different coupling constants $(c_3,c_4)$.}
\vspace{-8mm}
\end{center}
\end{figure}

We extend further our study and employ other Skyrme EDFs to test whether our conclusions are EDF-independent or not. These calculations are done for the $^{100,132}$Sn isotopes with the SGII, SIII, SkM* and SLy5 interactions~\cite{CHABANAT1997710,CHABANAT1998231}. On top of the EDF-dependence, we also check the LEC-dependence of the results by adopting three different sets.  First, we study the GT strength B(GT) of the double-magic nucleus $^{100}$Sn. Since the energy window for $\beta$-decay is as high as 7.4 MeV~\cite{AUDI2003337}, most of the GT strength is accessible by $\beta$-decay. Fig.~\hyperref[energy]{\ref*{100Sn-132Sn}(a)} shows the calculated GT strength B(GT) for the $\beta$-decay of $^{100}$Sn to the dominant $J^\pi = 1^{+}$ states in the daughter nucleus $^{100}$In. The experimental data from two groups~\cite{WOS:000305466800032,WOS:000282433500005} are shown with the bars of double arrows. The theoretical results come both from RPA and RPA + PVC models. In the figure,  open (filled) symbols represent the results with the OBC only (with  both the OBC and TBC). For the results with TBC, symbols with blue, green and purple colors correspond to the results obtained by adopting the coupling constants $(c_3,c_4)$ = $(-3.2,5.4)$ \cite{PhysRevC.68.041001}, $(-4.78,3.96)$ \cite{PhysRevC.67.044001} and $(-3.4,3.4)$ \cite{EPELBAUM2005362} in GeV$^{-1}$, respectively, while the short-range coupling constant is assigned as $\bar{c}_D=0$ for all the cases~\cite{PhysRevC.105.034349}. It is shown that the results with only the OBC always overestimate the experimental data. The results with OBC+TBC show good agreement with the experimental data of Ref. \cite{WOS:000305466800032} but are larger than the value of Ref. \cite{WOS:000282433500005}. Results given by the RPA + PVC model, regardless if only OBC or also TBC are included, are systematically lower than the corresponding results in the RPA model with the same Skyrme interactions. This illustrates that  the effect of many-body correlations beyond RPA model and TBC are both important in explaining the quenching phenomena.

The last two rows in Fig.~\hyperref[energy]{\ref*{100Sn-132Sn}(a)} show the results calculated by the {\it ab initio} coupled-cluster theory with (filled symbols) and without (open symbols) the contribution of TBC~\cite{WOS:000466716100012}. In these {\it ab initio}  calculations, two different chiral nuclear forces,  the 1.8/2.0 (EM) interaction~\cite{PhysRevC.83.031301} and the NN-N$^4$LO+3N$_{\mathrm{lnl}}$ interaction~\cite{WOS:000466716100012}, were adopted. These calculations show quantitatively lower B(GT) values than those of the present calculations,  while the quenching produced by TBC is similar to our case.

Fig.~\hyperref[energy]{\ref*{100Sn-132Sn}(b)} presents the theoretical and experimental results for the $\beta$-decay of $^{132}$Sn. In the RPA framework, all four Skyrme interactions predict this nucleus to be stable, i.e., no GT strength within the $Q_\beta$ window. In the PVC framework, the SGII interaction yields no GT strength inside the $Q_\beta$ window. Instead, the parameter sets SIII, SkM*, and SLy5 create strength below the $Q_\beta$ value. Compared with the experimental value in   Ref.~\cite{PhysRevC.51.500},  the RPA+PVC model with SIII and SkM*, combined with TBC,  provides a reasonable estimate of the nuclear $\beta$-decay matrix element of $^{132}$Sn, while SLy5 overestimates the B(GT) value by a factor of about 6.

\begin{table}[htbp]
    \centering
    \caption{Contributions of the four terms in Eq. (\ref{NMEs}) to the B(GT) value and its quenching.}
    \label{terms}
    \begin{ruledtabular}
    \begin{tabular}{rccccc} \\[-10pt]
        PVC & $\sum \text{B}_{\text{GT}}^1$ & $\sum \text{B}_{\text{GT}}^{1,3}$ & $\sum \text{B}_{\text{GT}}^{1,3,4}$ & Exp. & \textbf{$q$} \\ [2pt]
        \midrule
        $^{56}$\! Ni  & 0.212 & 0.133 & 0.137 & 0.152 & 0.80 \\ [2pt]
        $^{100}$\! Sn & 15.168 & 8.923 & 8.930 & $9.17_{-2.37}^{+2.64}$ & 0.77 \\ [2pt]
        $^{132}$\! Sn & 1.892 & 1.025 & 1.021 & $0.43_{-0.05}^{+0.06}$ & 0.73 \\
        \midrule
        RPA & $\sum \text{B}_{\text{GT}}^1$ & $\sum \text{B}_{\text{GT}}^{1,3}$  & - & Exp. & \textbf{$q$}  \\
        \midrule
        $^{56}$\! Ni  & 0      & 0     &  - & 0.152                 & -    \\ [2pt]
        $^{100}$\! Sn & 16.007 & 9.412 & -  &$9.17_{-2.37}^{+2.64}$ & 0.77 \\ [2pt]
        $^{132}$\! Sn & 0      & 0     & -  &$0.43_{-0.05}^{+0.06}$ & -
    \end{tabular}
    \end{ruledtabular}
\end{table}

To clarify the role of each contribution of the terms in Eq. (\ref{NMEs}) to the total B(GT) value, Table \ref{terms} shows the contributions of the four terms in Eq. (\ref{NMEs}) to the B(GT) in the RPA and RPA + PVC calculations by using the SkM* interaction together with the TBC of LECs, $(c_3,c_4)$ = $(-4.78,3.96)$ in GeV$^{-1}$ \cite{PhysRevC.67.044001} and $\bar{c}_D=0$. The quantity $\sum \text{B}_{\text{GT}}^i$ in Table \ref{terms} is obtained by only considering the contribution of $i$-th term $(i=1\sim4)$  in Eq. (\ref{NMEs}) for the GT matrix elements. By definition, the fourth term is absent in the RPA calculations. The second term in Eq. (\ref{NMEs}) remain identically zero because the one-body operator cannot mediate transitions from the ground state to the $2p$-$2h$ components of the excited-state wavefunction. We find that the contributions of TBC come predominantly from the third term, while the fourth term has only a minimal influence. Physically, the third term represents the effect of TBC on the $1p$-$1h$ component, and the fourth term represents the effect of TBC on the $1p$-$1h$ $\otimes$ phonon component. We quantify the TBC effect by defining the quenching factor $q$,
\begin{eqnarray}
    q = \sqrt{\frac{\sum \text{B}_{\text{GT}}^{1,3,4}}{\sum \text{B}_{\text{GT}}^{1}}}.\label{quen}
\end{eqnarray}
The values of $q$ in Table \ref{terms} are the extracted quenching factors from the results given by RPA and RPA + PVC with TBC for $^{56}$Ni, $^{100}$Sn, and $^{132}$Sn, respectively. The obtained quenching factors are in the range of $0.73-0.80$, aligning closely with the commonly adopted empirical quenching factor $q\approx 0.75$ \cite{TOWNER1987263}. For $^{56}$Ni and $^{132}$Sn, no B(GT) value is observed within the $Q_\beta$ window in the RPA calculation, and consequently no quenching factor $q$ is available.

To demonstrate the roles of momentum-dependent and short-range terms of the TBC on the GT strength, we calculate the contributions of these terms for $^{100}$Sn. The SkM* interaction and three $(c_3,c_4)$ parameter sets are used for this aim. Table \ref{momentum} shows the B(GT) with and without momentum-dependent term. It is shown that the contribution of the momentum-dependent term remains negligible regardless of the adopted LECs.

\begin{table}[!]
\centering
\caption{The contribution of momentum-dependent term (MDT) of chiral TBC in $^{100}$Sn. The results are obtained with three different $(c_3,c_4)$ parameter sets. }
\label{momentum}
\begin{ruledtabular}
\begin{tabular}{lcc}  \\[-10pt]
\textbf{$(c_3,c_4)$} & {w/o MDT} & {with MDT} \\ [2pt]
\hline
\\[-7pt]
$(-3.2, 5.4)$   & 8.270 & 8.290 \\
$(-4.78, 3.96)$ & 8.930 & 8.950 \\
$(-3.4, 3.4)$   & 9.948 & 9.970 \\
\end{tabular}
\end{ruledtabular}
\end{table}

\begin{figure}[!]
\begin{center}
  \includegraphics[width=0.40\textwidth]{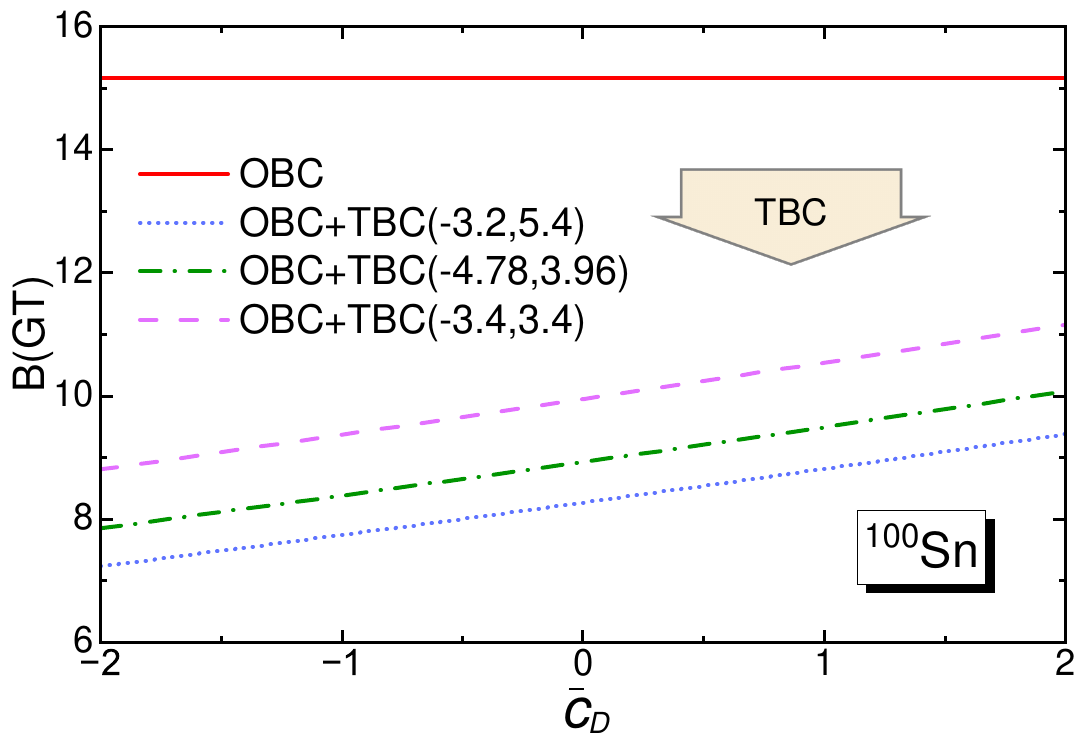}
  \caption{\label{cD}(Color online). Evolution of the B(GT) in $^{100}$Sn as a function of the coupling constant $\bar{c}_D$ for different groups of the ($c_3$, $c_4$) coupling constants.}
\vspace{-8mm}
\end{center}
\end{figure}

We  study also the effect of the short-range term  in Eq. \eqref{TBC-operator}, take $^{100}$Sn as an example.  Fig. \ref{cD} shows the calculated B(GT) as a function of the short-range coupling constant $\bar{c}_D$, where $\bar{c}_D$ changes from -2 to 2 together with three different $c_3$ and $c_4$ parameter sets.  It is found that the B(GT) increase linearly with increasing $\bar{c}_D$ for all three LEC sets, which is  similar to the results of Refs.~\cite{PhysRevC.98.031301, WOS:000466716100012}. We find that the change in B(GT) is about 2 unit for the change of $\bar{c}_D$  value from -2 to  2.  This change of B(GT)  is about 20\% to 30\% of the total contributions of the TBC.

\textit{Summary}-- We have studied the quenching of the axial-vector current coupling constant $g_A$ in $\beta$-decay, by taking into account the chiral TBC contributions on top of RPA+PVC model, which  includes the most relevant many-body correlations for the description of nuclear responses. The calculations are performed in the doubly-closed shell nuclei $^{56}$Ni, $^{100}$Sn  and $^{132}$Sn, adopting the microscopic wavefunctions obtained by the RPA+PVC model with Skyrme EDFs. We found that the effect of TBC hinders substantially the $\beta$-decay matrix elements in all the three nuclei, and produces better agreement with the experimental observations, irrespective to the adopted Skyrme EDFs in the RPA+PVC calculations. The quenching factor $q\approx0.73-0.80$, extracted as the combined effect of PVC+TBC, is almost identical to the phenomenological value $q\approx 0.75$,  which is commonly used in shell model and RPA calculations. Among the several contributions of chiral TBC, the meson-exchange terms give the major contribution to the quenching, to the choice of LEC parameters, while the momentum-dependent term has a negligible contribution. We examined also the short-range term contribution and found that it has a smaller effect  than the major TBC contributions. In the future, we will study the effects of chiral TBC on the nuclear matrix elements in two-neutrino double-$\beta$ decay and neutrinoless double-$\beta$ decay by using the same state-of-the-art model based on the EDF theory.

\textit{Acknowledgments}-- B.-L. Wang would like to thank H. Zhou and L.-J. Wang for helpful discussions. G. Col\`o thanks J. Mene\'ndez for very useful discussions. This work is partly supported by the National
Natural Science Foundation of China under Grant Nos. 12275025, 11975096, 12405135, 12135004, the National Key R$\&$D Program of China under Grant No. 2023YFA1606401, and the Fundamental Research Funds for the Central Universities under Grant No. 2020NTST06. H.S. is supported by Chinese Academy of Sciences CAS President's International Fellowship Initiative (PIFI) Grant No. 2024PVA0003-Y1 and the JSPS Grant-in-Aid for Scientific Research (C) under Grant No. JP26K07079.

\bibliography{magnetic_weak_process} 
\bibliographystyle{apsrev4-1} %

\end{document}


\title{Supplemental Material for ``The quenching of the axial-vector coupling constant $g_A$ in $\beta$-decay: joint effects from chiral two-body currents and many-body correlations''}

\author{Bin-Lei Wang}
\affiliation{Key Laboratory of Beam Technology of Ministry
of Education, School of Physics and Astronomy, Beijing
Normal University, Beijing 100875, China}

\author{Wan-Li Lv}
\affiliation{School of Nuclear Science and Technology, Lanzhou University, Lanzhou 730000, China}
\affiliation{Frontiers Science Center for Rare isotopes, Lanzhou University, Lanzhou 730000, China}

\author{Li-Gang Cao}
\affiliation{Key Laboratory of Beam Technology of Ministry
of Education, School of Physics and Astronomy, Beijing
Normal University, Beijing 100875, China}
\affiliation{Key Laboratory of Beam Technology of Ministry
of Education, Institute of Radiation Technology, Beijing
Academy of Science and Technology, Beijing 100875, China}

\author{Yi-Fei Niu}
\affiliation{School of Nuclear Science and Technology, Lanzhou University, Lanzhou 730000, China}
\affiliation{Frontiers Science Center for Rare isotopes, Lanzhou University, Lanzhou 730000, China}

\author{Gianluca Col\`o}
\affiliation{Dipartimento di Fisica, Universit\`a degli Studi di Milano, and INFN, Sezione di Milano, Via Celoria 16, 20133 Milano, Italy}

\author{Hiroyuki Sagawa}
\affiliation{RIKEN Nishina Center, Wako 351-0198, Japan}
\affiliation{Center for Mathematics and Physics, University of Aizu, Aizu-Wakamatsu 965-8560, Japan}
\affiliation{Institute of Theoretical Physics, Chinese Academy of Sciences, Beijing 100190, China}

\author{Feng-Shou Zhang}
\affiliation{Key Laboratory of Beam Technology of Ministry
of Education, School of Physics and Astronomy, Beijing
Normal University, Beijing 100875, China}
\affiliation{Key Laboratory of Beam Technology of Ministry
of Education, Institute of Radiation Technology, Beijing
Academy of Science and Technology, Beijing 100875, China}
\affiliation{Center of Theoretical Nuclear Physics, National Laboratory of Heavy Ion Accelerator of Lanzhou, Lanzhou 730000, China}

\date{\today}


\maketitle



This supplemental material includes mainly the theoretical framework employed in the paper and the numerical check of our results, which is presented in the following five sections: (I) Gamow-Teller (GT) matrix elements in $\beta$-decay with the two-body currents, (II) the basic formalism of the RPA + PVC model, (III) the Gamow-Teller matrix elements in RPA + PVC model with chiral two-body currents, (IV) the expansion of the transition matrix elements with spherical harmonic oscillator basis, and (V) the numerical tests of the results.

\section{Gamow-Teller Matrix Elements in $\beta$-decay}
The form of the atomic transition matrix element in $\beta$-decay is \cite{walecka1975muon},
\begin{eqnarray}
    \left\langle F\left| H_{\beta}\right|I\right\rangle = -\frac{G_{\beta}}{\sqrt{2}} \,l^{\,\mu} \int d^{3} x \left\langle f \left| e^{-i \bm{q} \cdot \boldsymbol{x}} \mathbb{J}_{\mu }(\boldsymbol{x}) \right|i \right\rangle, \label{atomic}
\end{eqnarray}
where $H_\beta$ denotes the Hamiltonian density, $\bm{q}$ is the momentum transferred to leptons and $l^{\,\mu}$ is the leptonic matrix element that depends only on initial and final lepton wave functions, $\mathbb{J}_{\mu}(\boldsymbol{x})$ is the nuclear weak
current. The term, $\left\langle f \left| \mathbb{J}_{\mu}(\boldsymbol{x}) \right| i \right\rangle$, is the matrix element of the nuclear current operator between initial $|i\rangle$ and final $|f\rangle$ nuclear states.

Under the long-wavelength approximation, the Gamow-Teller (GT) matrix elements in even-even nuclei are given by,
\begin{eqnarray}
    \left\langle 1_\omega^+\left|\mathbb{M}^\lambda\right|0^+\right\rangle = \int d \boldsymbol{x}\left\langle 1_\omega^+\left| e^{-i \bm{q} \cdot \boldsymbol{x}} \mathbb{J}_{\mu}(\boldsymbol{x})\right|0^+\right\rangle \approx \int d \boldsymbol{x}\left\langle 1_\omega^+\left| \mathbb{J}_{\mu}(\boldsymbol{x})\right|0^+\right\rangle.
\end{eqnarray}
In the standard model, the nuclear current is a sum of vector and axial-vector parts,
\begin{eqnarray}
  \mathbb{J}_{\mu}(\boldsymbol{x}) = \mathbb{J}_{\mu}^V(\boldsymbol{x}) + \mathbb{J}_{\mu}^A(\boldsymbol{x}).
\end{eqnarray}
Here, we consider only the axial-vector current in the limit of zero momentum transfer \cite{PhysRevC.105.034349}. The total nuclear current operator takes the form,
\begin{eqnarray}
  \mathbb{J}_{\mu}^A(\boldsymbol{x}) = \mathbb{J}_{1b}^A(\boldsymbol{x}) + \mathbb{J}_{2b}^A(\boldsymbol{x}) + \cdots.
\end{eqnarray}

\section{RPA+PVC model}
For the GT$^-$ transition, the coupled form of the RPA excitation operator $\mathcal{Q}_\omega^\dagger$ takes the form,
\begin{eqnarray}
    && \mathcal{Q}_{\omega}^\dagger(JM) = \sum_{{\pi}{\bar{\nu}}} X_{{\pi}{\bar{\nu}}}^{\omega J} \mathcal{A}_{{\pi}{\bar{\nu}}}^\dagger(JM) - \sum_{{\nu}{\bar{\pi}}} Y_{{\nu}{\bar{\pi}}}^{\omega J} \tilde{\mathcal{A}}_{{\nu}{\bar{\pi}}}(JM),
\end{eqnarray}
where $\pi$ and $\bar{\nu}$ represent a proton in a particle state and a neutron in a hole state, respectively. $\mathcal{A}_{{\pi}{\bar{\nu}}}^\dagger(JM)$ represents the creation operator of the one particle-one hole (1$p$-1$h$) configuration and $\tilde{\mathcal{A}}_{{\nu}{\bar{\pi}}}(JM)$ represents the time-reversed state of its Hermitian conjugate, taking the following form,
\begin{eqnarray}
    && \mathcal{A}_{\pi\bar{\nu}}^\dagger(JM) = \sum_{m_\pi m_{\bar{\nu}}} (-)^{j_{\bar{\nu}}-m_{\bar{\nu}}} C_{j_\pi m_\pi j_{\bar{\nu}} -m_{\bar{\nu}}}^{JM} c_{\pi}^\dagger c_{\bar{\nu}}, \\
    && \tilde{\mathcal{A}}_{\nu\bar{\pi}}(JM) \equiv (-)^{J+M} \mathcal{A}_{\nu\bar{\pi}}(J-M) = \sum_{m_\nu m_{\bar{\pi}}} (-)^{J+M+j_{\bar{\pi}}-m_{\bar{\pi}}} C_{j_\nu m_\nu j_{\bar{\pi}} -m_{\bar{\pi}}}^{J-M} c_{\bar{\pi}}^\dagger c_{\nu}.
\end{eqnarray}

The coupled form of the RPA+PVC excitation operator $\mathcal{O}_{\omega}^\dagger$ is,
\begin{eqnarray}
    \mathcal{O}_\omega^\dagger(JM) = \mathcal{O}_{1\omega}^\dagger(JM) + \mathcal{O}_{2\omega}^\dagger(JM) && = \sum_{{\pi}{\bar{\nu}}} X_{{\pi}{\bar{\nu}}}^{\omega J} \mathcal{A}_{{\pi}{\bar{\nu}}}^\dagger(JM) - \sum_{{\nu}{\bar{\pi}}} Y_{{\nu}{\bar{\pi}}}^{\omega J} \tilde{\mathcal{A}}_{{\nu}{\bar{\pi}}}(JM) \notag\\
    && + \sum_{{\pi}{\bar{\nu}} n J_{\pi\bar{\nu}} J^\prime} X_{{\pi}{\bar{\nu}} n}^{\omega J} \mathcal{B}_{{\pi}{\bar{\nu}} J_{\pi\bar{\nu}} n J^\prime}^\dagger(JM) - \sum_{{\nu}{\bar{\pi}} n J_{\nu\bar{\pi}} J^\prime} Y_{{\nu}{\bar{\pi}} n}^{\omega J} \tilde{\mathcal{B}}_{{\nu}{\bar{\pi}} J_{\pi\bar{\nu}} n J^\prime}(JM),
\end{eqnarray}
where $\omega$ represents the $\omega$-th 1$^{+}$ excited state of the nucleus, $J$ represents the nuclear spin, $M$ represents its third component, $J_{\pi\bar{\nu}}$ represents the angular momentum of the 1$p$-1$h$ configuration, $J^\prime$ represents the phonon angular momentum, and $n$ represents the $n$-th phonon state. $\mathcal{B}_{{\pi}{\bar{\nu}} J_{\pi\bar{\nu}} n J^\prime}^\dagger(JM)$ represents the creation operator of the 1$p$-1$h$ $\otimes$ phonon configuration and $\tilde{\mathcal{B}}_{{\nu}{\bar{\pi}} J_{\pi\bar{\nu}} n J^\prime}(JM)$ represents the time-reversed state of its Hermitian conjugate, taking the following form,
\begin{eqnarray}
    && \mathcal{B}_{{\pi}{\bar{\nu}} J_{\pi\bar{\nu}} n J^\prime}^\dagger(JM) = \sum_{M_{\pi\bar{\nu}}M^\prime}  C_{J_{{\pi\bar{\nu}}} M_{{\pi\bar{\nu}}} J^\prime M^\prime}^{JM} \mathcal{A}_{\pi\bar{\nu}}^\dagger(J_{\pi\bar{\nu}}M_{\pi\bar{\nu}}) \mathcal{Q}_{n}^\dagger(J^\prime M^\prime), \\
    && \tilde{\mathcal{B}}_{{\nu}{\bar{\pi}} J_{\pi\bar{\nu}} n J^\prime}(JM) \equiv (-)^{J+M} \mathcal{B}_{{\nu}{\bar{\pi}} J_{\pi\bar{\nu}} n J^\prime}(J-M) = (-)^{J+M} \sum_{M_{\nu\bar{\pi}}M^\prime}  C_{J_{{\nu\bar{\pi}}} M_{{\nu\bar{\pi}}} J^\prime M^\prime}^{J-M} \mathcal{Q}_{n}(J^\prime M^\prime) \mathcal{A}_{\nu\bar{\pi}}(J_{\nu\bar{\pi}}M_{\nu\bar{\pi}}).
\end{eqnarray}

With the operator defined above, we get the RPA + PVC equation,
\begin{eqnarray}
  \left(\begin{array}{cccc}
        A_{\pi \bar{\nu},\pi^\prime \bar{\nu}^\prime} & B_{\pi\bar{\nu},\nu^\prime \bar{\pi}^\prime} & A_{\pi\bar{\nu},\pi^\prime \bar{\nu}^\prime n^\prime} & 0 \\
        -B^\ast_{\nu\bar{\pi},\pi^\prime \bar{\nu}^\prime} & -A^\ast_{\nu\bar{\pi},\nu^\prime \bar{\pi}^\prime} & 0 & -A^\ast_{\nu\bar{\pi},\nu^\prime \bar{\pi}^\prime n^\prime} \\
        A_{\pi\bar{\nu}n,\pi^\prime \bar{\nu}^\prime} & 0 & A_{\pi\bar{\nu}n,\pi^\prime \bar{\nu}^\prime n^\prime} & 0 \\
        0 & -A^\ast_{\nu \bar{\pi} n,\nu^\prime \bar{\pi}^\prime} & 0 & -A^\ast_{\nu \bar{\pi} n,\nu^\prime \bar{\pi}^\prime n^\prime}
    \end{array}\right)
    \left(\begin{array}{ccc}
        X^\omega_{\pi^\prime \bar{\nu}^\prime}   \\
        Y^\omega_{\nu^\prime \bar{\pi}^\prime}   \\
        X^\omega_{\pi^\prime \bar{\nu}^\prime n^\prime}   \\
        Y^\omega_{\nu^\prime \bar{\pi}^\prime n^\prime}
    \end{array}\right)
    = E_\omega
        \left(\begin{array}{ccc}
        X^\omega_{\pi \bar{\nu}}   \\
        Y^\omega_{\nu \bar{\pi}}   \\
        X^\omega_{\pi \bar{\nu} n}   \\
        Y^\omega_{\nu \bar{\pi} n}
        \end{array}\right). \label{RPAPVC}
\end{eqnarray}
where $E_\omega$ represents the excitation energy of the $\omega$-th 1$^{+}$ excited state.

\section{GT Matrix Elements in RPA+PVC model}

The GT strength B(GT) is proportional to the square of matrix element $\left\langle 1_\omega^+\left| \mathbb{M}^{\lambda} \right| 0^+\right\rangle$, which can be obtained within the RPA+PVC approach. The second-quantized forms of the one-body current operator $\mathbb{M}^{\lambda}_{1b}$ and the two-body current operator $\mathbb{M}^{\lambda}_{2b}$ are given as follows,
\begin{gather}
    \mathbb{M}_{1b}^{\lambda} = \sum_{ab}\left\langle a\left|\mathcal{M}_{1b}^{\lambda}\right|b\right\rangle c_a^\dagger c_b, \\
    \mathbb{M}_{2b}^{\lambda} = \frac{1}{2}\sum_{aba^\prime b^\prime}\left\langle aa^\prime\left|\mathcal{M}_{2b}^{\lambda}\right|bb^\prime\right\rangle c_a^\dagger c_{a^\prime}^\dagger c_{b^\prime} c_{b},
\end{gather}
and,
\begin{eqnarray}
    &&\left\langle 1_\omega^+\left| \mathbb{M}^{\lambda} \right| 0^+\right\rangle = \left\langle 1_\omega^+\left| \mathbb{M}^{\lambda} \right| {\rm PVC}\right\rangle = \left\langle {\rm PVC}\left|\mathcal{O}_\omega \mathbb{M}^{\lambda} \right| {\rm PVC}\right\rangle \notag \\
    &=& \left\langle {\rm PVC}\left|\left[\mathcal{O}_\omega,\mathbb{M}^{\lambda}\right]\right| {\rm PVC}\right\rangle \overset{\rm QBA}{\approx} \left\langle {\rm HF}\left|\left[\mathcal{O}_\omega,\mathbb{M}^{\lambda}\right]\right| {\rm HF}\right\rangle = \left\langle {\rm HF}\left|\left[\mathcal{O}_{1\omega} + \mathcal{O}_{2\omega}, \mathbb{M}_{1b}^{\lambda}+\mathbb{M}_{2b}^{\lambda}\right]\right| {\rm HF}\right\rangle \notag\\
    &=& \left\langle {\rm HF}\left|\left[\mathcal{O}_{1\omega},\mathbb{M}_{1b}^{\lambda}\right]\right| {\rm HF}\right\rangle + \left\langle {\rm HF}\left|\left[\mathcal{O}_{2\omega},\mathbb{M}_{1b}^{\lambda}\right]\right| {\rm HF}\right\rangle + \left\langle {\rm HF}\left|\left[\mathcal{O}_{1\omega},\mathbb{M}_{2b}^{\lambda}\right]\right| {\rm HF}\right\rangle + \left\langle {\rm HF}\left|\left[\mathcal{O}_{2\omega},\mathbb{M}_{2b}^{\lambda}\right]\right| {\rm HF}\right\rangle.  \label{4term}
\end{eqnarray}

The first matrix elements read,
\begin{eqnarray}
    \left\langle {\rm HF}\left|\left[\mathcal{O}_{1\omega},\mathbb{M}_{1b}^{\lambda}\right]\right| {\rm HF}\right\rangle = \sum_{\pi\bar{\nu}} X_{\pi\bar{\nu}}^{\omega J \ast} \frac{1}{\hat{J}} \left\langle \pi \left\|\mathcal{M}_{1b}^{\lambda}\right\|\bar{\nu}\right\rangle + \sum_{\nu\bar{\pi}} Y_{\nu\bar{\pi}}^{\omega J \ast} (-)^{J+j_{\bar{\pi}}+j_\nu+1} \frac{1}{\hat{J}} \left\langle \bar{\pi} \left\|\mathcal{M}_{1b}^{\lambda}\right\|\nu\right\rangle, \label{term1}
\end{eqnarray}
where $\left|\pi\right\rangle$ represents the single-particle state in the $J$-scheme with good quantum numbers $n_\pi$, $l_\pi$, $j_\pi$, and $\tau_\pi$, and $\hat{J}=\sqrt{2J+1}$.

The second matrix elements read,
\begin{eqnarray}
    \left\langle {\rm HF}\left|\left[\mathcal{O}_{2\omega},\mathbb{M}_{1b}^{\lambda}\right]\right| {\rm HF}\right\rangle = 0. \label{term2}
\end{eqnarray}
The second matrix elements in Eq. (\ref{term2}) remain identically zero because the one-body operator can only act on a single nucleon at a time and therefore cannot mediate transitions from the ground state to the 2p-2h components of the excited-state wave function.

The third matrix elements read,
\begin{eqnarray}
    && \left\langle {\rm HF}\left|\left[\mathcal{O}_{1\omega},\mathbb{M}_{2b}^{\lambda}\right]\right| {\rm HF}\right\rangle = \sum_{\pi\bar{\nu}a} X_{\pi\bar{\nu}}^{\omega J \ast} \sum_{J_{\pi a} J_{\bar{\nu}a}} (-)^{J_{\bar{\nu}a}+J+j_a+j_\pi} \frac{\hat{J}_{\bar{\nu}a}\hat{J}_{\pi a}}{\hat{J}}  \left\{\begin{array}{lll} j_{\bar{\nu}} & j_a & J_{\bar{\nu}a} \\ J_{\pi a} & J & j_\pi \end{array}\right\} \left\langle \pi a \left\|^{J_{\pi a}}\mathcal{M}_{2b}^{\lambda} \right\|\bar{\nu} a \right\rangle^{J_{\bar{\nu}a}}_A \notag\\
    && + \sum_{\nu\bar{\pi}a} Y_{\nu\bar{\pi}}^{\omega J \ast} \sum_{J_{\nu a} J_{\bar{\pi}a}} (-)^{J_{\nu a}+j_a+j_\nu} \frac{\hat{J}_{\bar{\pi}a}\hat{J}_{\nu a}}{\hat{J}}  \left\{\begin{array}{lll} j_{\bar{\pi}} & j_a & J_{\bar{\pi}a} \\ J_{\nu a} & J & j_\nu \end{array}\right\}  \left\langle \bar{\pi}a \left\|^{J_{\bar{\pi}a}} \mathcal{M}_{2b}^{\lambda} \right\|\nu a \right\rangle^{J_{\nu a}}_A,  \label{term3}
\end{eqnarray}
where $\left\langle \pi a \left\|^{J_{\pi a}}\mathcal{M}_{2b}^{\lambda} \right\|\bar{\nu} a \right\rangle^{J_{\bar{\nu}a}}_A = \left\langle \pi a \left\|^{J_{\pi a}}\mathcal{M}_{2b}^{\lambda} \right\|\bar{\nu} a \right\rangle^{J_{\bar{\nu}a}} - (-)^{j_{\bar{\nu}}+j_a+J_{\bar{\nu}a}} \left\langle \pi a \left\|^{J_{\pi a}}\mathcal{M}_{2b}^{\lambda} \right\|a \bar{\nu} \right\rangle^{J_{\bar{\nu}a}}$.

The fourth matrix elements read,
\begin{eqnarray}
    && \left\langle {\rm HF}\left|\left[\mathcal{O}_{2\omega},\mathbb{M}_{2b}^{\lambda}\right]\right| {\rm HF}\right\rangle = \sum_{\pi \bar{\nu} n J_{\pi \bar{\nu}} J^\prime} X_{\pi \bar{\nu} n}^{\omega J\ast} \sum_{p^\prime h^\prime} X_{p^\prime h^\prime}^{n J^\prime \ast} \sum_{J_{\pi p^\prime} J_{\bar{\nu} h^\prime}} \frac{\hat{J}_{\pi \bar{\nu}} \hat{J}^\prime \hat{J}_{\bar{\nu} h^\prime} \hat{J}_{\pi p^\prime}}{\hat{J}}  \left\{\begin{array}{lll} j_\pi & j_{p^\prime} & J_{\pi p^\prime} \\ j_{\bar{\nu}} & j_{h^\prime} & J_{\bar{\nu} h^\prime} \\ \!J_{\pi \bar{\nu}} & J^\prime & \,\,J \end{array}\right\} \left\langle \pi p^\prime \left\|^{J_{\pi p^\prime}} \mathcal{M}_{2b}^{\lambda} \right\|\bar{\nu} h^\prime \right\rangle^{J_{\bar{\nu} h^\prime}}_A \notag\\
    && + \sum_{\nu \bar{\pi} n J_{\nu \bar{\pi}} J^\prime}Y_{\nu \bar{\pi} n}^{\omega J\ast} \sum_{p^\prime h^\prime} X_{p^\prime h^\prime}^{n J^\prime \ast} \sum_{J_{\nu p^\prime} J_{\bar{\pi} h^\prime}} (-)^{J_{\bar{\pi} h^\prime}-J_{\nu p^\prime}+J} \frac{\hat{J}_{\nu \bar{\pi}} \hat{J}^\prime\hat{J}_{\bar{\pi} h^\prime}\hat{J}_{\nu p^\prime}}{\hat{J}}  \left\{\begin{array}{lll} j_\nu & j_{p^\prime} & J_{\nu p^\prime} \\ j_{\bar{\pi}} & j_{h^\prime} & J_{\bar{\pi} h^\prime} \\ \!J_{\nu \bar{\pi}} & J^\prime & \,\,J \end{array}\right\}  \left\langle \bar{\pi} h^\prime \left\|^{J_{\bar{\pi}} h^\prime} \mathcal{M}_{2b}^{\lambda} \right\|\nu p^\prime \right\rangle^{J_{\nu p^\prime}}_A,  \label{term4}
\end{eqnarray}
in Eq. (\ref{term4}), only the $X_{p^\prime h^\prime}^{n J^\prime \ast}$ part of the phonon operator is appeared,  while the $Y_{p^\prime h^\prime}^{n J^\prime \ast}$ part is vanished because the contraction.

Three types of $\beta$-decay exist, $\beta^-$, $\beta^+$, and electron capture (EC), and the corresponding decay windows are defined as follows.

For $\beta^-$ decay, the decay window reads,
\begin{eqnarray}
    E_{{\rm RPA}} \leq m_\nu-m_\pi-m_e \approx 0.782 ({\rm MeV}),
\end{eqnarray}
where $E_{RPA}$ represents the excitation energy of the daughter nucleus relative to the parent nucleus.

For $\beta^+$, the decay window reads,
\begin{eqnarray}
    E_{{\rm RPA}} \leq m_\pi-m_\nu-m_e \approx -1.804 ({\rm MeV}),
\end{eqnarray}

For EC decay, the decay window reads,
\begin{eqnarray}
    E_{{\rm RPA}} \leq m_\pi-m_\nu+m_e-B_i \approx -0.782 - B_i \,({\rm MeV}),
\end{eqnarray}
where $B_i$ represents the electron binding energy in the atom and $i$ represents the electron shell.

\section{Transition Matrix Elements Within the Harmonic Oscillator Basis }

The single-particle wavefunctions in the reduced matrix elements of Eq. (\ref{term1},\ref{term3},\ref{term4}) are obtained from the Skyrme-Hartree Fock mean field calculation and are expanded in the spherical harmonic oscillator basis,
\begin{eqnarray}
    \psi_j(\bmr) = \sum_{i} \int_{\bmr^\prime} \phi_i^\dagger(\bmr^{\prime}) \psi_j(\bmr^{\prime}) \phi_i(\bmr) = \sum_{i} c_i^j \phi_i(\bmr),
\end{eqnarray}
where $\psi_j$ represents the single-particle wave function from the Hartree-Fock method, $\phi_i$ represents the spherical harmonic oscillator wave function, and $c_i^j$ represents the basis expansion coefficient.

When the wavefunctions are expanded in the spherical harmonic oscillator basis, the one-body and two-body matrix elements are expressed as the followings,
\begin{eqnarray}
    \left\langle a \left\| \mathcal{M}_{1b}^{\lambda} \right\| b \right\rangle &=& \left\langle n_a l_a j_a \tau_a \left\| \mathcal{M}_{1b}^{\lambda} \right\| n_b l_b j_b \tau_{b} \right\rangle \notag \\
    &=& \sum_{n_1 n_2} c_{n_1}^{n_{a}} c_{n_2}^{n_{b}} \left\langle n_1 l_a j_a \tau_a \left\| \mathcal{M}_{1b}^{\lambda} \right\| n_2 l_b j_b \tau_{b} \right\rangle,\\
    \left\langle a a^\prime \left\|^{J_{aa^\prime }} \mathcal{M}_{2b}^{\lambda} \right\| b b^\prime \right\rangle^{J_{bb^\prime }} &=& \left\langle n_a l_a j_a \tau_a, n_{a^\prime } l_{a^\prime } j_{a^\prime } \tau_{a^\prime } \left\|^{J_{aa^\prime }} \mathcal{M}_{2b}^{\lambda} \right\| n_b l_b j_b \tau_{b}, n_{b^\prime } l_{b^\prime } j_{b^\prime } \tau_{b^\prime } \right\rangle^{J_{bb^\prime }} \notag\\
     &=& \sum_{n_1 n_2 n_3 n_4} c_{n_1}^{n_{a}} c_{n_2}^{n_{a^\prime }} c_{n_3}^{n_{b}} c_{n_4}^{n_{b^\prime }} \left\langle n_1 l_a j_a \tau_a, n_2 l_{a^\prime } j_{a^\prime } \tau_{a^\prime } \left\|^{J_{aa^\prime }} \mathcal{M}_{2b}^{\lambda} \right\| n_3 l_b j_b \tau_{b}, n_4 l_{b^\prime } j_{b^\prime } \tau_{b^\prime } \right\rangle^{J_{bb^\prime }}.
\end{eqnarray}

The two-body matrix element involves relative coordinates, which requires to transform the spatial coordinate of the single-particle wavefunctions from the absolute coordinate system to the Jacobi coordinate system by the Moshinsky transformation, and the transformation method is described as below \cite{KAMUNTAVICIUS2001191},
\begin{eqnarray}
    \left|n_{1} l_{1}, n_{2} l_{2}\right\rangle^{\lambda \mu} = \sum_{n l N L} \left|N L, n l \right\rangle^{\lambda \mu} \left\langle\left\langle N L, n l \big| n_{1} l_{1}, n_{2} l_{2} \right\rangle\right\rangle^{\lambda},  \label{MOSHINSKY}
\end{eqnarray}
where $\left\langle\left\langle N L, n l \big| n_{1} l_{1}, n_{2} l_{2} \right\rangle\right\rangle^{\lambda}$ represents the harmonic-oscillator brackets (HOBs) and the transformation satisfies the following relation \cite{KAMUNTAVICIUS2001191},

\begin{align}
    2 n_{1}+l_{1}+2 n_{2}+l_{2} = 2 n+l+2 N+L.  \label{MOSHINSKY-1}
\end{align}

\section{Numerical Check}

\begin{figure}
\begin{center}
  \includegraphics[width=0.500\textwidth]{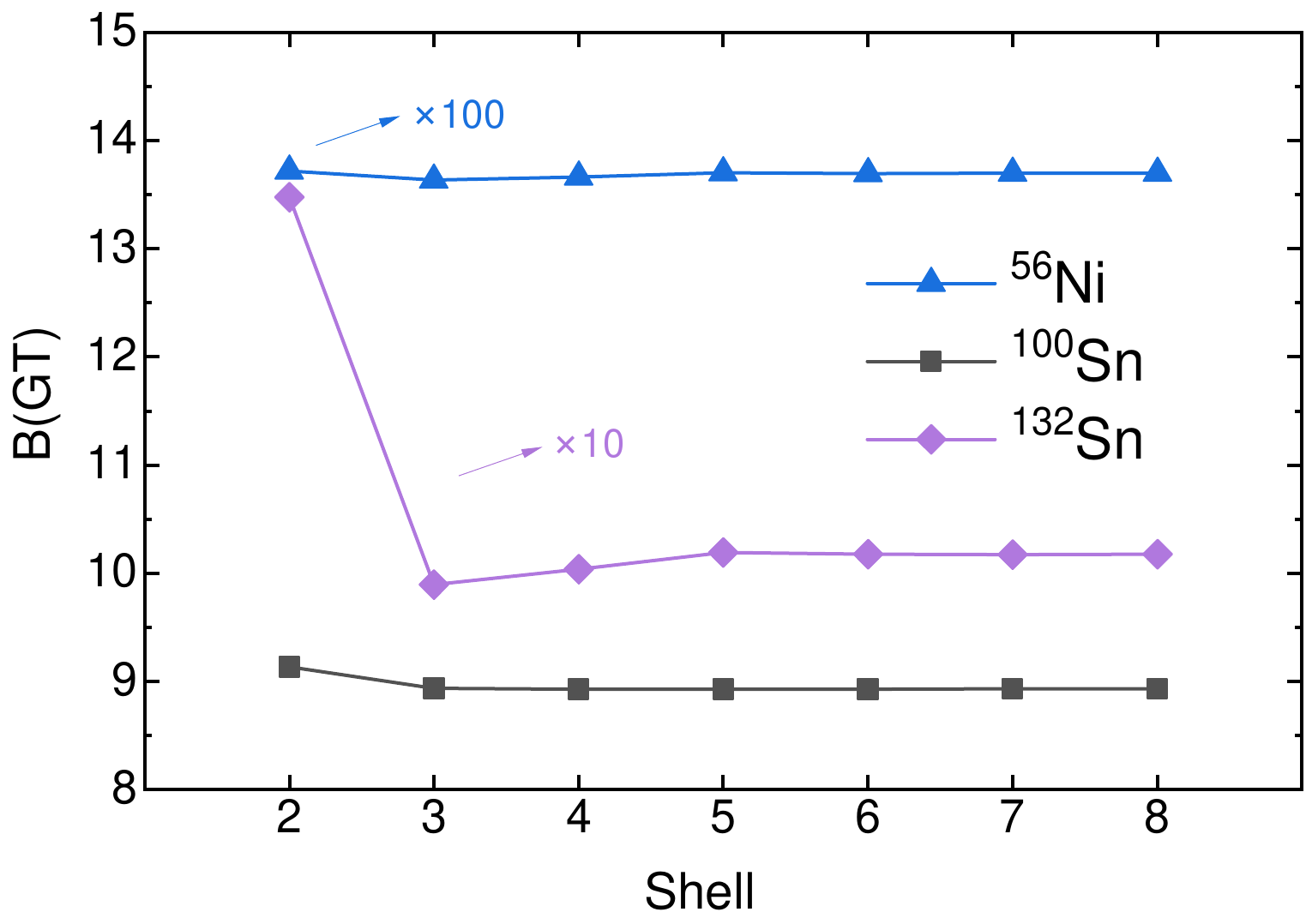}
  \caption{\label{shellall}(Color online). The convergence of the calculated B(GT) for $^{56}$Ni, $^{100}$Sn and $^{132}$Sn as a function of the shell number used in the harmonic oscillator basis expansion. The data for $^{132}$Sn ($^{56}$Ni) shown in the figure are the values obtained by multiplying 10 (100) on its real values.}
\end{center}
\end{figure}

Fig. \ref{shellall} shows the convergence of the calculated B(GT) in $\beta$-decay for $^{56}$Ni, $^{100}$Sn and $^{132}$Sn under different shell choice in the expansion of the wave function. The results are obtained within RPA+PVC calculations with SkM$^*$ interaction. The data for $^{100}$Sn in the figure are the real values, while the data for $^{132}$Sn ($^{56}$Ni) shown in the figure are the values obtained by multiplying 10 (100) on its real values. The results demonstrate that 6 to 8 oscillator shells are sufficient for the convergence in these doubly magic nuclei. 

\begin{table}
\centering
\caption{
The main contributions of particle-hole configurations and the phonon couplings to the matrix elements in Eq. (\ref{4term}) for the RPA+PVC state at 4.451 MeV in $^{100}$Sn, the subscripts $\pi$ and $\nu$ refer to proton and neutron states, respectively.}
\label{momentum}
\begin{ruledtabular}
\begin{tabular}{cccc}  \\[-10pt]
1$p$-1$h$ conf. & ${|X_{\nu\bar{\pi}}|}^2 / {|Y_{\pi\bar{\nu}}|}^2$ & Term (1) & Term (3) \\ [2pt]
\hline
\\[-7pt]
$[\pi1g_{9/2}^{-1},\nu1g_{7/2}]$ & $ 9.321\times 10^{-1}$ & $ 2.340\times 10^{ 0}$ & $-5.229\times 10^{-1}$  \\[1pt]
$[\pi1g_{7/2},\nu1g_{9/2}^{-1}]$ & $ 1.829\times 10^{-3}$ & $-1.031\times 10^{-1}$ & $ 2.291\times 10^{-2}$  \\[1pt]
$[\pi1d_{5/2}^{-1},\nu2d_{3/2}]$ & $ 4.280\times 10^{-4}$ & $-4.369\times 10^{-3}$ & $-1.731\times 10^{-3}$  \\[1pt]
$[\pi1d_{3/2}^{-1},\nu2d_{5/2}]$ & $ 5.350\times 10^{-4}$ & $ 2.095\times 10^{-3}$ & $-3.819\times 10^{-3}$  \\[1pt]
$[\pi1f_{5/2}^{-1},\nu2f_{7/2}]$ & $ 8.252\times 10^{-4}$ & $ 1.339\times 10^{-3}$ & $-5.105\times 10^{-3}$  \\[1pt]
$[\pi2s_{1/2}^{-1},\nu3s_{1/2}]$ & $ 2.357\times 10^{-4}$ & $-1.046\times 10^{-3}$ & $-1.950\times 10^{-3}$  \\[1pt]
$[\pi1d_{5/2}^{-1},\nu2d_{5/2}]$ & $ 1.986\times 10^{-4}$ & $-8.090\times 10^{-4}$ & $-2.166\times 10^{-3}$  \\[1pt]
$[\pi1g_{9/2}^{-1},\nu4g_{7/2}]$ & $ 3.107\times 10^{-4}$ & $ 7.674\times 10^{-4}$ & $ 1.883\times 10^{-3}$  \\[3pt]
\hline  \\[-9pt]
1$p$-1$h$ $\otimes$ phonon conf. & ${|X_{\nu\bar{\pi}n}|}^2 / {|Y_{\pi\bar{\nu}n}|}^2$ & Term (4) \\ [2pt]
\hline
\\[-7pt]
$[\pi1f_{7/2}^{-1},\nu1g_{7/2}]^1\otimes 1_6^-$ & $6.946\times 10^{-4}$ & $-3.104\times 10^{-4}$ \\[1pt]
$[\pi1f_{7/2}^{-1},\nu1g_{7/2}]^1\otimes 1_4^-$ & $7.705\times 10^{-4}$ & $ 2.824\times 10^{-4}$ \\[1pt]
$[\pi1g_{9/2}^{-1},\nu2d_{5/2}]^2\otimes 2_1^+$ & $2.908\times 10^{-4}$ & $ 2.803\times 10^{-4}$ \\[1pt]
$[\pi1g_{9/2}^{-1},\nu2d_{3/2}]^3\otimes 2_2^+$ & $7.109\times 10^{-3}$ & $ 2.331\times 10^{-4}$ \\[1pt]
$[\pi1g_{9/2}^{-1},\nu1h_{9/2}]^1\otimes 1_2^-$ & $2.020\times 10^{-4}$ & $ 2.049\times 10^{-4}$ \\[1pt]
$[\pi1g_{9/2}^{-1},\nu2d_{5/2}]^3\otimes 2_1^+$ & $1.807\times 10^{-3}$ & $-2.042\times 10^{-4}$ \\[1pt]
$[\pi1f_{7/2}^{-1},\nu1g_{7/2}]^1\otimes 1_2^-$ & $1.542\times 10^{-4}$ & $ 1.765\times 10^{-4}$ \\[1pt]
$[\pi1g_{9/2}^{-1},\nu2d_{5/2}]^3\otimes 2_2^+$ & $1.222\times 10^{-2}$ & $ 1.756\times 10^{-4}$ \label{conf}
\end{tabular}
\end{ruledtabular}
\end{table}

Table \ref{conf} presents the main contributions from particle-hole configurations and the phonon couplings to the matrix elements in Eq. (\ref{4term}) for the $1^+$ state at 4.451 MeV in $^{100}$Sn obtained from RPA+PVC calculations with both the one and two-body currents, the results are obtained with SkM$^*$ interaction. It is shown that Term(1) yields the dominant contribution, Term(3) gives a moderate contribution, and the contributions from Term(4) are the smallest one. 

\begin{figure*}[!]
\begin{center}
  \includegraphics[width=1.000\textwidth]{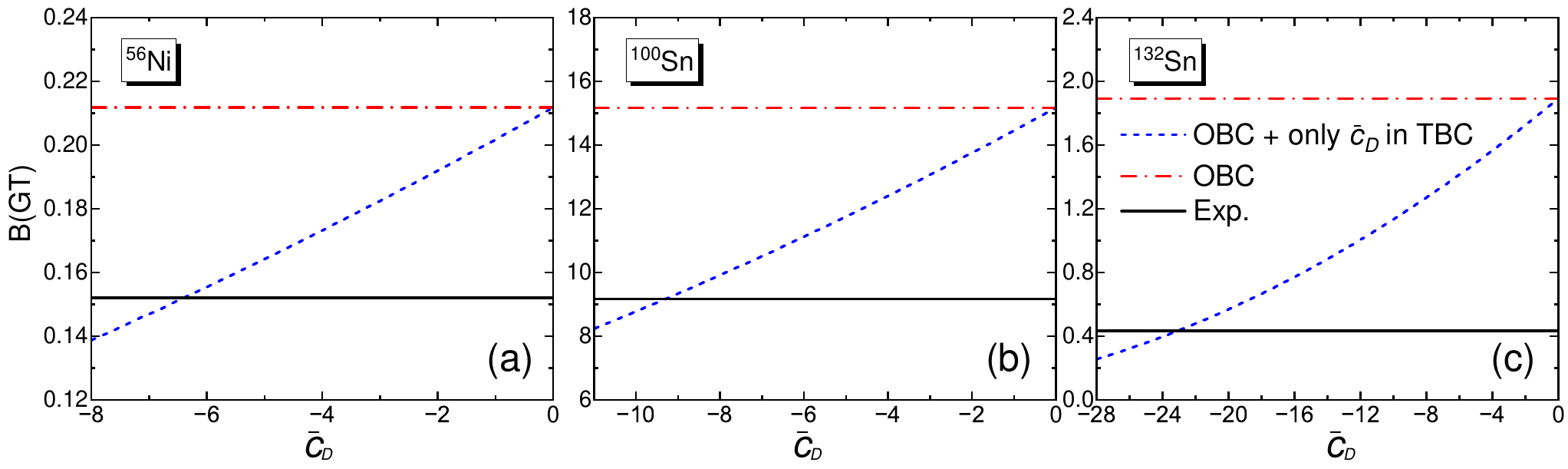}
  \caption{\label{only-cD}(Color online). B(GT) values in $\beta$-decay of $^{56}$Ni, $^{100}$Sn, and $^{132}$Sn as a function of the short-range coupling constant $\bar{c}_D$, the calculations are done within RPA+PVC considering only the short-range term contribution in two-body currents (blue dashed lines). The red dash-dotted lines show the results from only the one-body current, and the black solid lines denote the experimental values\cite{nndc}.}
\end{center}
\end{figure*}

Since the effective nucleon-nucleon interaction we adopted is the zero-range Skyrme interaction, we have also investigated the contribution of only short-range $\bar{c}_D$ term for $^{56}$Ni, $^{100}$Sn, and $^{132}$Sn, this means we block the contributions from the finite-range $c_3$ and $c_4$ terms as well as the momentum-dependent term in TBC. In this case we need to adjust the parameter $\bar{c}_D$ to fit the experimental B(GT) values. The calculations are done with SkM* interaction. Fig. \ref{only-cD} displays the variations of the B(GT) values for $^{56}$Ni, $^{100}$Sn, and $^{132}$Sn with respect to $\bar{c}_D$ when only the $\bar{c}_D$ term is included. One observes that B(GT) decreases consistently as $\bar{c}_D$ decreases. To reproduce the experimental values, the adjusted $\bar{c}_D$ parameters are -6.4 for $^{56}$Ni, -9.3 for $^{100}$Sn, and -23.1 for $^{132}$Sn. The values of $\bar{c}_D$ are very different for these three nuclei because the short-range contribution is relatively small compared to the finite-range terms, thus more extra $\bar{c}_D$ values are needed to sufficiently suppress the B(GT) strengths in certain nuclei.

\bibliography{magnetic_weak_process} 

\bibliographystyle{apsrev4-1} %